\documentclass[nofootinbib,twocolumn,showpacs,preprintnumbers,pre,aps]{revtex4-1}

\usepackage[dvipdfmx]{graphicx}
\usepackage{bm}
\usepackage{amsmath}
\usepackage{amssymb}
\usepackage{color}

\begin{document}

\title{Time-correlation functions for odd Langevin systems}

\author{Kento Yasuda}\email{yasudak@kurims.kyoto-u.ac.jp}

\affiliation{
Research Institute for Mathematical Sciences, 
Kyoto University, Kyoto 606-8502, Japan}

\author{Kenta Ishimoto}

\affiliation{
Research Institute for Mathematical Sciences, 
Kyoto University, Kyoto 606-8502, Japan}

\author{Akira Kobayashi}

\affiliation{
Department of Chemistry, Graduate School of Science,
Tokyo Metropolitan University, Tokyo 192-0397, Japan}

\author{Li-Shing Lin}

\affiliation{
Department of Chemistry, Graduate School of Science,
Tokyo Metropolitan University, Tokyo 192-0397, Japan}

\author{Isamu Sou}

\affiliation{
Department of Chemistry, Graduate School of Science,
Tokyo Metropolitan University, Tokyo 192-0397, Japan}

\author{Yuto Hosaka}

\affiliation{
Max Planck Institute for Dynamics and Self-Organization (MPI DS), 
Am Fassberg 17, 37077 G\"{o}ttingen, Germany}

\author{Shigeyuki Komura}\email{komura@wiucas.ac.cn}

\affiliation{
Wenzhou Institute, University of Chinese Academy of Sciences, 
Wenzhou, Zhejiang 325001, China} 

\affiliation{
Oujiang Laboratory, Wenzhou, Zhejiang 325000, China}

\affiliation{
Department of Chemistry, Graduate School of Science,
Tokyo Metropolitan University, Tokyo 192-0397, Japan}


\begin{abstract}
We investigate the statistical properties of fluctuations in active systems that are governed by non-symmetric
responses. 
Both an underdamped Langevin system with an odd resistance tensor and an overdamped Langevin system with 
an odd elastic tensor are studied. 
For a system in thermal equilibrium, the time-correlation functions should satisfy time-reversal symmetry and the 
anti-symmetric parts of the correlation functions should vanish. 
For the odd Langevin systems, however, we find that the anti-symmetric parts of the time-correlation 
functions can exist and that they are proportional to either the odd resistance coefficient or the odd elastic constant.  
This means that the time-reversal invariance of the correlation functions is broken due to the presence of odd 
responses in active systems. 
Using the short-time asymptotic expressions of the time-correlation functions, one can estimate an odd elastic 
constant of an active material such as an enzyme or a motor protein. 
\end{abstract}

\maketitle

\section{Introduction}
\label{Sec:Int}

Over the last decades, various active systems such as motor proteins, bacteria, flocks of birds
and fishes were intensively studied as fundamental problems of non-equilibrium statistical mechanics 
and biophysics~\cite{Gompper20}.
Recently, investigations have begun to characterize these active systems with non-symmetric response functions 
such as odd viscosity or odd elasticity~\cite{Banerjee17,Scheibner20,Hosaka22,Vitelli22}.
In thermal equilibrium, response functions such as resistance coefficient and elastic modulus need 
to satisfy certain symmetry properties.    
For example, the resistance coefficient tensor of a rigid object in a viscous fluid should be a symmetric matrix 
owing to time-reversal symmetry of low-Reynolds-number hydrodynamic fluid~\cite{KuboBook,DoiBook}.
Such a special property is one example of more general Onsager's reciprocal relations that 
restrict the symmetry of transport coefficient matrices within the linear response theory.
For non-equilibrium active systems, however, such reciprocal relations are often violated, and the 
response functions generally consist of both symmetric (even) and anti-symmetric (odd) parts.

One of the non-reciprocal responses that have been investigated is the odd viscosity proposed 
by Avron some years ago~\cite{Avron98}.
In general, viscosity is a fourth-rank tensor that linearly connects a stress tensor and a rate-of-strain tensor.
According to Onsager's reciprocal theorem, the viscosity tensor should be symmetric for ordinary passive fluids
under the exchange of the pairs of the indices.
For an active suspension of rotary motor, however, such symmetry is violated and an odd part of the viscosity 
can exist~\cite{Banerjee17,Markovich21}.
The microscopic origin of odd viscosity is attributed to the broken time-reversal symmetry of the constituent 
elements.
Recently, several people derived the generalized Green-Kubo relation for odd viscosity that arises when the 
time-reversal symmetry of stress fluctuation is violated~\cite{Epstein20,Hargus20,Han21}.
Some of the present authors calculated the resistance tensor of a two-dimensional (2D) liquid domain immersed in a fluid 
with odd viscosity and showed that it has non-zero anti-symmetric components~\cite{Hosaka21,Hosaka21b}.

Recently, Scheibner \textit{et al.}\ introduced the concept of odd elasticity to describe non-conserved 
interactions in active materials~\cite{Scheibner20}.
An elastic modulus is, in general, a fourth-rank tensor tensor that linearly connects stress and strain tensors.
For a passive system, it should be symmetric under the exchange of the pairs of the indices because elastic 
forces are generally conservative~\cite{LandauBook}.
For active systems, on the other hand, an elastic modulus can have both even and odd 
parts~\cite{Scheibner20,Braverman21}.
Importantly, the active moduli quantify the amount of work extracted along quasistatic strain cycles.
In our recent paper, we used the variational principle of the Onsager-Machlup integral to describe 
the stochastic dynamics of a micromachine with odd elasticity~\cite{Yasuda22}.
Furthermore, the concept of odd viscoelasticity~\cite{Banerjee21} and odd diffusion tensor have also 
been reported~\cite{Hargus21}.

In this paper, we investigate the statistical properties of fluctuations in active systems that are governed by 
non-symmetric responses. 
We employ linear Langevin equations with odd responses and obtain various time-correlation 
functions~\cite{Weiss03,Weiss07,Hosaka17,Sou19,Sou21}.
Generally, symmetry properties of time-correlation functions can be discussed in terms of 
their transformation under time-translational and time-reversal operations~\cite{KuboBook,DoiBook}.
When the system is in a steady state, the correlation functions need to satisfy the time-translational invariance.
Although the time-reversal invariance of the cross-correlation functions is also satisfied in a thermal
equilibrium situation~\cite{DoiBook}, we show that such symmetry is violated in the 
presence of non-symmetric responses. 
Two limiting cases will be investigated in detail; linear Langevin systems with a non-symmetric 
resistance tensor and a non-symmetric elastic tensor. 
In both cases, the time-reversal symmetry of correlation functions is violated in the presence 
of odd responses.  
We show that one can estimate an odd elastic constant of an active material such as an enzyme or 
a motor protein by using the short-time asymptotic expressions of the time-correlation functions.
As an example of this application, we demonstrate the estimation of odd elasticity from 
the numerical simulation of our enzyme model with chemical reactions~\cite{Yasuda21a}.

In the next section, we briefly summarize the symmetry properties of time-correlation functions
both in equilibrium and out-of-equilibrium. 
In Sec.~\ref{Sec:Under}, we investigate the linear Langevin system with an odd resistance tensor.
In Sec.~\ref{Sec:Over}, we discuss the Langevin system with an odd elastic tensor.
In Sec.~\ref{Sec:enzyme}, we explain the application of our results to the model 
enzyme system.
A summary and some further discussion are given in Sec.~\ref{Sec:Dis}.

\section{Time-correlation functions}
\label{Sec:SymTCF}

Let us introduce $N$-dimensional position variables $x_\alpha(t)$ ($\alpha=1,2,\cdots,N$) 
and velocity variables $v_\alpha(t)=\dot x_\alpha(t)$, where the dot indicates the time derivative.
The variables $x_\alpha(t)$ represent, for example, positions of colloid particles
in a suspension or structural parameters of a protein molecule.
Considering only the fluctuations, 
we assume that the averages of $x_\alpha(t)$ and $v_\alpha(t)$ vanish, i.e., 
$\langle x_\alpha(t)\rangle=0$ and $\langle v_\alpha(t)\rangle=0$, where 
$\langle \cdots \rangle$ indicates the ensemble average.
Notice that $x_\alpha(t)$ is even and $v_\alpha(t)$ is odd under time-reversal transformation.

Let us define the following position-position, position-velocity, and velocity-velocity time-correlation matrices:
\begin{align}
\phi_{\alpha\beta}(t)&=\langle x_\alpha(t)x_\beta(0)\rangle, \\
\chi_{\alpha\beta}(t)&=\langle v_\alpha(t)x_\beta(0)\rangle, \\
\psi_{\alpha\beta}(t)&=\langle v_\alpha(t)v_\beta(0)\rangle.
\end{align}
The equal-time-correlation functions for $t=0$ are defined with a bar such as 
$\bar\phi_{\alpha\beta}=\phi_{\alpha\beta}(t=0)$ and we similarly define 
$\bar\chi_{\alpha\beta}$ and $\bar\psi_{\alpha\beta}$.

Generally, one can decompose the time-correlation matrices into symmetric and 
anti-symmetric parts as 
\begin{align}
\phi_{\alpha\beta}(t)& =\phi_{\alpha\beta}^\mathrm S(t)+\phi_{\alpha\beta}^\mathrm A(t),
\\
\chi_{\alpha\beta}(t)& =\chi_{\alpha\beta}^\mathrm S(t)+\chi_{\alpha\beta}^\mathrm A(t),
\\
\psi_{\alpha\beta}(t)& =\psi_{\alpha\beta}^\mathrm S(t)+\psi_{\alpha\beta}^\mathrm A(t),
\end{align}
where, for example, $\phi_{\alpha\beta}^\mathrm S(t)=\phi_{\beta\alpha}^\mathrm S(t)$ and 
$\phi_{\alpha\beta}^\mathrm A(t)=-\phi_{\beta\alpha}^\mathrm A(t)$ hold.
Notice that the mathematical meanings of the superscripts ``$\mathrm S$" and 
``$\mathrm A$" are the same as ``even" and ``odd", respectively.
However, we employ ``$\mathrm S$" and ``$\mathrm A$" for time-correlation functions, whereas 
``even" and ``odd" are used for viscosity and elasticity due to convention~\cite{Vitelli22}.
In the following, we argue the properties of the above time-correlation functions when time-translational 
invariance and time-reversal invariance are satisfied.

\subsection{Time-translational invariance}

If the system is in a steady state (both in equilibrium and out-of-equilibrium), time-correlation 
functions do not depend on the origin of time and satisfy the relation 
$\langle a(t)b(t')\rangle=\langle a(t-t')b(0)\rangle$~\cite{KuboBook,DoiBook}.
As a result of such time-translational invariance, we have 
\begin{align}
\langle a(t)b(0)\rangle=\langle a(0)b(-t)\rangle=\langle b(-t)a(0)\rangle.
\end{align}
Hence the position-position and velocity-velocity correlation matrices satisfy the following relations:
\begin{align}
\phi_{\alpha\beta}(t)&=\phi_{\beta\alpha}(-t),
\label{TTSphi}
\\
\psi_{\alpha\beta}(t)&=\psi_{\beta\alpha}(-t).
\label{TTSpsi}
\end{align}
Concerning the position-velocity correlation function, we have 
$\chi_{\alpha\beta}(t)=\dot\phi_{\alpha\beta}(t)$ (and also 
$\psi_{\alpha\beta}(t)=-\dot\chi_{\alpha\beta}(t)$).
Hence the time-translational invariance requires 
\begin{align}
\chi_{\alpha\beta}(t)=-\chi_{\beta\alpha}(-t).
\label{TTSchi}
\end{align}

The above symmetry properties in the steady state can be conveniently expressed in terms of the 
symmetric and anti-symmetric parts of the correlation functions as 
\begin{align}
& \phi_{\alpha\beta}^\mathrm S(t)=\phi_{\alpha\beta}^\mathrm S(-t),
\hspace{12pt}
\phi_{\alpha\beta}^\mathrm A(t)=-\phi_{\alpha\beta}^\mathrm A(-t),
\label{TTSphi-EO}\\
& \chi_{\alpha\beta}^\mathrm S(t)=-\chi_{\alpha\beta}^\mathrm S(-t),
\hspace{12pt}
\chi_{\alpha\beta}^\mathrm A(t)=\chi_{\alpha\beta}^\mathrm A(-t),
\label{TTSchi-EO}\\
& \psi_{\alpha\beta}^\mathrm S(t)=\psi_{\alpha\beta}^\mathrm S(-t),
\hspace{12pt}
\psi_{\alpha\beta}^\mathrm A(t)=-\psi_{\alpha\beta}^\mathrm A(-t).
\label{TTSpsi-EO}
\end{align}
In other words, $\phi_{\alpha\beta}^\mathrm S(t)$ and $\psi_{\alpha\beta}^\mathrm S(t)$ are even 
functions of time, while $\phi_{\alpha\beta}^\mathrm A(t)$ and $\psi_{\alpha\beta}^\mathrm A(t)$
are odd functions~\cite{KuboBook}.
On the other hand, $\chi_{\alpha\beta}^\mathrm S(t)$ and $\chi_{\alpha\beta}^\mathrm A(t)$
are odd and even functions of time, respectively.

For the equal-time-correlation functions, we set $t=0$ in Eqs.~(\ref{TTSphi}), (\ref{TTSchi}), and 
(\ref{TTSpsi}) 
\begin{align}
\bar\phi_{\alpha\beta}=\bar\phi_{\beta\alpha},\hspace{12pt}
\bar\chi_{\alpha\beta}=-\bar\chi_{\beta\alpha},\hspace{12pt}
\bar\psi_{\alpha\beta}=\bar\psi_{\beta\alpha}.
\label{TTS-Same}
\end{align}
Hence, $\bar\phi_{\alpha\beta}$ and $\bar\psi_{\alpha\beta}$ are symmetric matrices with 
respect to the exchange of the indices, while $\bar\chi_{\alpha\beta}$ is a completely anti-symmetric 
matrix.

\subsection{Time-reversal invariance in thermal equilibrium}

Next, we discuss the properties of the correlation functions in thermal equilibrium when time-reversal 
invariance holds~\cite{KuboBook,DoiBook}. 
In this situation, the correlation functions satisfy the relation 
$\langle a(t)b(0)\rangle_{\rm eq}=\varepsilon_a\varepsilon_b\langle a(-t)b(0)\rangle_{\rm eq}$, 
where $\varepsilon_{a(b)}$ takes the value $1$ or $-1$ depending on the time-reversal 
symmetry of the variable $a(b)$.  
For example, we have $\varepsilon_x=1$ and $\varepsilon_v=-1$, as mentioned before. 
Therefore, the equilibrium correlation functions need to satisfy the following symmetry relations:
\begin{align}
\phi_{\alpha\beta}^{\rm eq}(t)&=\phi_{\alpha\beta}^{\rm eq}(-t),
\label{TRSphi}\\
\chi_{\alpha\beta}^{\rm eq}(t)&=-\chi_{\alpha\beta}^{\rm eq}(-t),
\label{TRSchi}\\
\psi_{\alpha\beta}^{\rm eq}(t)&=\psi_{\alpha\beta}^{\rm eq}(-t).
\label{TRSpsi}
\end{align}

In thermal equilibrium, time-translational invariance is also satisfied and hence 
Eqs.~(\ref{TTSphi-EO})-(\ref{TTSpsi-EO}) hold simultaneously.
Then the anti-symmetric parts of the time-correlation matrices should vanish in equilibrium:
\begin{align}
\phi_{\alpha\beta}^{\rm A, eq}(t)=\chi_{\alpha\beta}^{\rm A, eq}(t)=\psi_{\alpha\beta}^{\rm A, eq}(t)=0.
\label{TRS-EO}
\end{align}
As a result, the time correlation matrices have only the symmetric parts that satisfy such as 
$\phi_{\alpha\beta}^{\rm S}(t)=\phi_{\beta\alpha}^{\rm S}(t)$.
Hence, the time-correlation matrices should be symmetric under the exchange of the two indices in 
thermal equilibrium.
In non-equilibrium situations, however, the anti-symmetric parts can exist because time-reversal invariance
can be violated. 
Moreover, we also have $\bar\chi_{\alpha\beta}^{\rm eq}=0$ by considering $t=0$ in Eq.~(\ref{TRSchi}).

\section{Underdamped Langevin system with odd resistance tensor}
\label{Sec:Under}

\begin{figure}[tb]
\begin{center}
\includegraphics[scale=0.35]{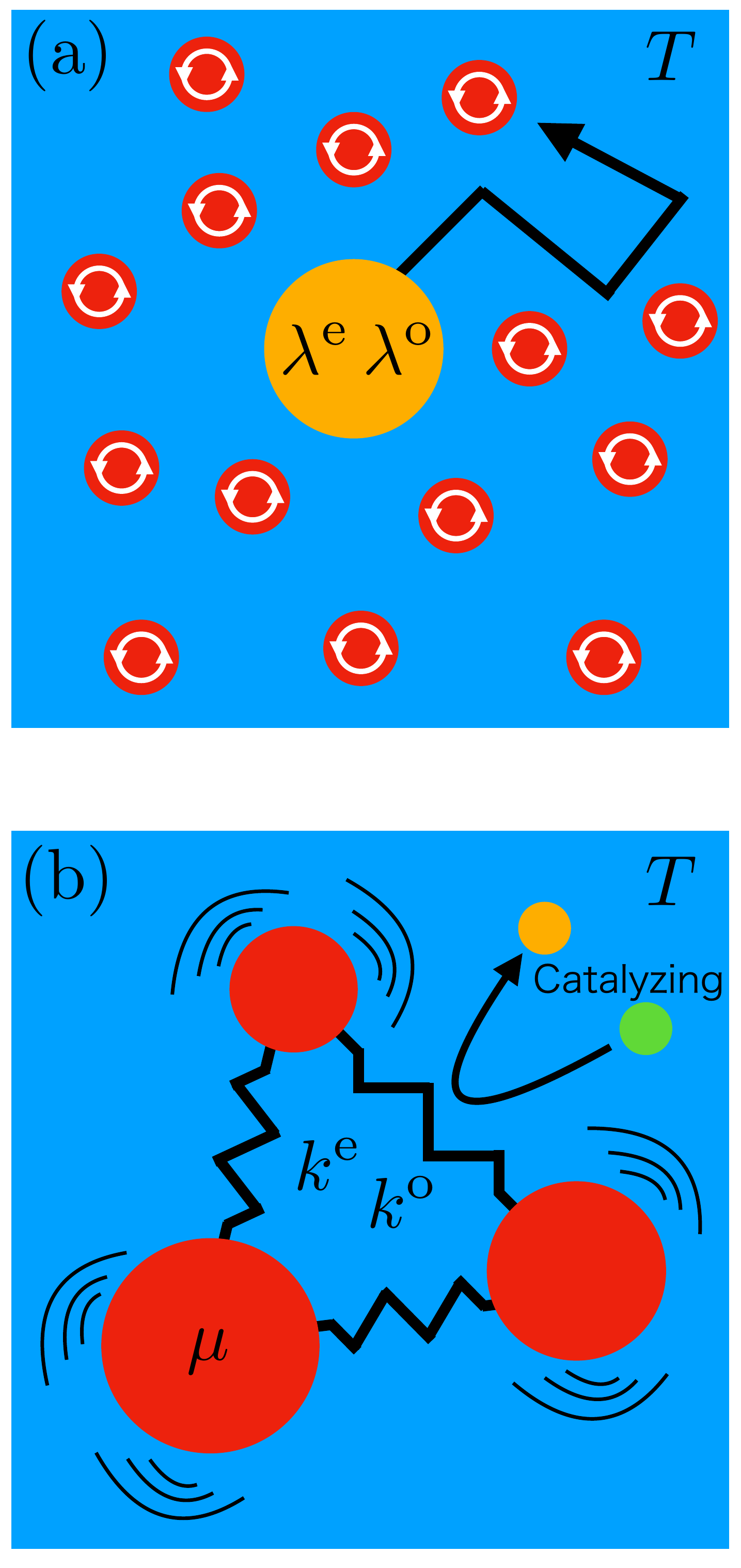}
\end{center}
\caption{
(a) Diffusion of a Brownian particle (orange circle) that experiences both even and odd resistance forces whose 
coefficients are given by $\lambda^\mathrm e$ and $\lambda^\mathrm o$, respectively. 
When the surrounding fluid is composed of active chiral elements (red circles), the resistance tensor can have 
both symmetric and anti-symmetric parts [see Eq.~(\ref{evenoddviscosity})].
(b) A coarse-grained model of an enzyme consisting of domains that are connected to springs.
A substrate (green circle) changes into a product (orange circle) via a catalytic chemical reaction.
Each spring is characterized by the even elastic constant $k^{\rm e}$ and the odd 
elastic constant $k^{\rm o}$ [see Eq.~(\ref{SymK})].
}
\label{model}
\end{figure}

\subsection{$N$ degrees of freedom}

In this section, we consider a free Brownian particle embedded in an active chiral fluid that is characterized 
by odd viscosity, as schematically shown in Fig.~\ref{model}(a).
In a two-dimensional space, the drag force acting on the particle due to the surrounding active fluid is given 
by a non-symmetric resistance tensor~\cite{Hosaka21,Hosaka21b}.
To describe the Brownian dynamics of the particle, we use the underdamped Langevin equation in the 
presence of a non-symmetric resistance tensor.

The underdamped linear Langevin equation for $N$ velocity variables $v_\alpha(t)$ can be written as~\cite{KuboBook,DoiBook,Weiss03,Weiss07}
\begin{align}
m\dot v_\alpha(t)=-\Lambda_{\alpha\beta}v_\beta(t)+mF_{\alpha\beta}\xi_\beta(t), 
\label{UnderLanEq}
\end{align}
where $m$ is the mass of a Brownian particle, $\Lambda_{\alpha\beta}$ is the resistance tensor~\cite{Hosaka21,Hosaka21b},
$F_{\alpha\beta}$ is the noise amplitude, and $\xi_\beta(t)$ is Gaussian white noise that satisfies  
\begin{align}
\langle\xi_\alpha(t)\rangle=0,\hspace{12pt}
\langle\xi_\alpha(t)\xi_\beta(t')\rangle=\delta_{\alpha\beta}\delta(t-t').
\label{WGN}
\end{align}
The strength of the noise can be conveniently characterized by the symmetric tensor defined 
by $B_{\alpha\beta}=F_{\alpha\gamma}F_{\beta\gamma}/2$.

For a passive system, the resistance tensor should be symmetric,  
$\Lambda_{\alpha\beta}=\Lambda_{\beta\alpha}$, due to Lorentz reciprocal theorem, or, more generally, 
Onsager's reciprocal relations~\cite{KuboBook,DoiBook}.
In addition, the second law of thermodynamics requires that it should be positive definite.
For an active system, however, $\Lambda_{\alpha\beta}$ can have an anti-symmetric part and 
we can generally decompose it as~\cite{Hosaka21,Hosaka21b}
\begin{align}
\Lambda_{\alpha\beta}=\Lambda_{\alpha\beta}^\mathrm e+\Lambda_{\alpha\beta}^\mathrm o.
\label{DC-o}
\end{align}
Here the symmetric (even) part and the anti-symmetric (odd) part satisfy 
$\Lambda_{\alpha\beta}^\mathrm e = \Lambda_{\beta\alpha}^\mathrm e$ and 
$\Lambda_{\alpha\beta}^\mathrm o = -\Lambda_{\beta\alpha}^\mathrm o$, respectively.
The linear Langevin equation in Eq.~(\ref{UnderLanEq}) can be analytically solved as described 
in Refs.~\cite{Weiss03,Weiss07} and also briefly summarized in Appendix~\ref{App:A}.

In thermal equilibrium, the equal-time velocity-velocity correlation function is determined by the equipartition 
theorem as~\cite{KuboBook,DoiBook}
\begin{align}
\bar \psi_{\alpha\beta} = \frac{k_\mathrm BT}{m}\delta_{\alpha\beta},
\label{equipartition}
\end{align}
where $k_{\rm B}$ is the Boltzmann constant and $T$ the temperature.
Then we solve the Lyapunov equation~\cite{Weiss03,Weiss07} in Eq.~(\ref{Lyapunov-eq}) for the noise strength $B_{\alpha\beta}$ as 
\begin{align}
B_{\alpha\beta}=\frac{k_\mathrm BT}{m^2}\Lambda_{\alpha\beta}^\mathrm e.
\end{align}
This is the fluctuation dissipation theorem for a passive system in which only 
$\Lambda_{\alpha\beta}^\mathrm e$ exists~\cite{KuboBook,DoiBook}.

Next we argue the velocity-velocity time-correlation matrix 
$\psi_{\alpha\beta}(t)=\langle v_\alpha(t)v_\beta(0)\rangle $.
For $N$ degrees of freedom, it is enough to know the short-time behavior of $\psi_{\alpha\beta}(t)$ to discuss its time-reversal symmetry.
As derived in Eqs.~(\ref{TCF-ST-e}) and (\ref{TCF-ST-o}) of Appendix~\ref{App:A}, we obtain the short-time behavior of 
$\psi_{\alpha\beta}(t)=
\psi_{\alpha\beta}^\mathrm S(t) + \psi_{\alpha\beta}^\mathrm A(t)$,
where the symmetric and anti-symmetric parts become 
\begin{align}
\psi_{\alpha\beta}^\mathrm S(t) & \approx\frac{k_\mathrm BT}{m}\left(\delta_{\alpha\beta}-\frac{\Lambda_{\alpha\beta}^\mathrm e|t|}{m}\right), 
\\
\psi_{\alpha\beta}^\mathrm A(t) & \approx-\frac{k_\mathrm BT}{m^2}\Lambda_{\alpha\beta}^\mathrm o t.
\label{Gen-ST-TCF-psi}
\end{align}
In the above, we have assumed that $|\bar \Lambda_i||t|/m\ll 1$ is satisfied for all the 
eigenvalues $\bar\Lambda_i$ of the matrix $\Lambda_{\alpha\beta}$.
In accordance with Eq.~(\ref{TTSpsi-EO}), $\psi_{\alpha\beta}^\mathrm S(t)$ is an even function of time 
and $\psi_{\alpha\beta}^\mathrm A(t)$ is an odd function.
In contrast to Eq.~(\ref{TRS-EO}), however, $\psi_{\alpha\beta}^\mathrm A$ does not vanish 
and time-reversal symmetry is broken when $\Lambda_{\alpha\beta}^\mathrm o \neq 0$.
This is an important consequence when the odd part of the resistance tensor exists.

\subsection{Two degrees of freedom}

\begin{figure}[tb]
\begin{center}
\includegraphics[scale=0.5]{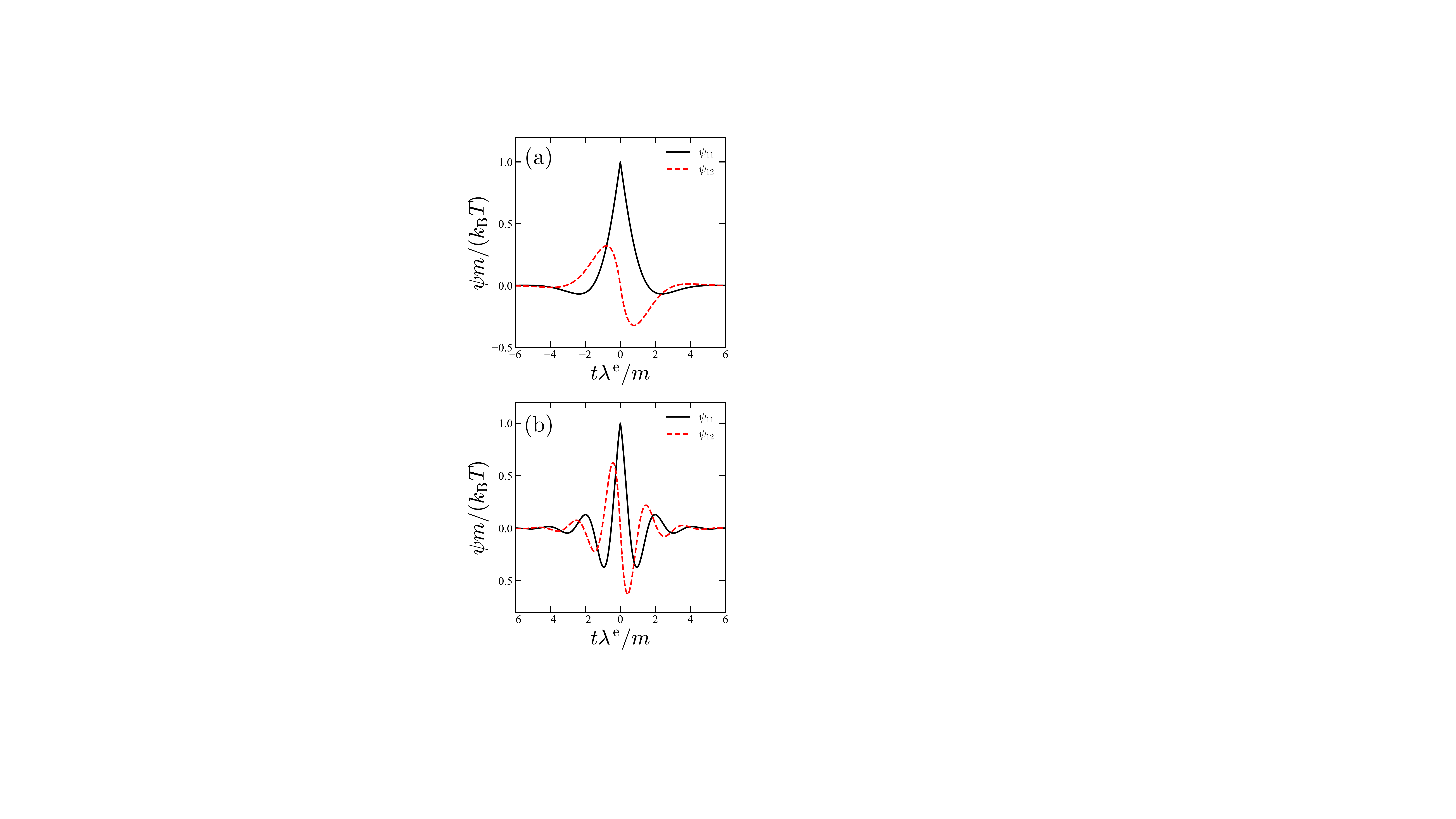}
\end{center}
\caption{
Plots of the scaled velocity-velocity correlation functions $\psi_{11}(t)$ (black solid line) and 
$\psi_{12}(t)$ (red dashed line) as a function of dimensionless time $\lambda^\mathrm e t/m$
when (a) $\lambda^\mathrm o/\lambda^\mathrm e=1$ and (b) $\lambda^\mathrm o/\lambda^\mathrm e=3$
for $N=2$ [see Eqs.~(\ref{2DTCF-psi-Sym}) and (\ref{2DTCF-psi-Asym})].
In both cases, $\psi_{11}(t)$ is an even function of time, while $\psi_{12}(t)$ is an odd function 
of time. 
Since $\psi_{12}(t)$ violates Eq.~(\ref{TRSpsi}), time-reversal invariance is broken 
in the presence of odd resistance tensor. 
In (b), we observe the oscillatory behavior of the correlation functions. 
Notice that the position-position correlation function $\phi_{\alpha\beta}$ behaves in the same 
way as $\psi_{\alpha\beta}$ as long as the proper scaling is made.
}
\label{Fig:TCF}
\end{figure}

To perform analytical calculations, we discuss the time-correlation matrix when $N=2$.
We further assume that the resistance tensor is given by the following form: 
\begin{align}
\Lambda_{\alpha\beta}=\lambda^\mathrm e \delta_{\alpha\beta}+\lambda^\mathrm o\epsilon_{\alpha\beta},
\label{evenoddviscosity}
\end{align}
where $\epsilon_{\alpha\beta}$ is the 2D Levi-Civita tensor with $\epsilon_{11}=\epsilon_{22}=0$ and 
$\epsilon_{12}=-\epsilon_{21}=1$.
Notice that $\lambda^\mathrm e>0$ while $\lambda^\mathrm o$ can take both positive and negative values.
More general cases are discussed in Appendix~\ref{App:B}.

The symmetric and anti-symmetric parts of $\psi_{\alpha\beta}(t)$ become
\begin{align}
\psi_{\alpha\beta}^\mathrm S(t) & = 
\frac{k_\mathrm BT}{m}e^{- \lambda^\mathrm e|t|/m} 
\cos(\lambda^\mathrm ot/m) \delta_{\alpha\beta}, 
\label{2DTCF-psi-Sym}
\\
\psi_{\alpha\beta}^\mathrm A(t) & =
- \frac{k_\mathrm BT}{m}e^{- \lambda^\mathrm e|t|/m} 
\sin(\lambda^\mathrm ot/m) \epsilon_{\alpha\beta}.
\label{2DTCF-psi-Asym}
\end{align}
We confirm again that $\psi_{\alpha\beta}^\mathrm S(t)$ is an even function of time and 
$\psi_{\alpha\beta}^\mathrm A(t)$ is an odd function.
Also, the existence of $\psi_{\alpha\beta}^\mathrm A$ when $\lambda^\mathrm o\ne 0$ indicates 
the violation of time-reversal symmetry.

In Fig~\ref{Fig:TCF}, we plot $\psi_{11}(t)$ (black solid line) and $\psi_{12}(t)$ (red dashed line)
as a function of dimensionless time $\lambda^\mathrm e t/m$.
The other parameter is $\lambda^\mathrm o/\lambda^\mathrm e=1$ in Fig.~\ref{Fig:TCF}(a) and 
$\lambda^\mathrm o/\lambda^\mathrm e=3$ in Fig.~\ref{Fig:TCF}(b).
In both cases, $\psi_{11}(t)$ is an even function, while $\psi_{12}(t)$ is an odd function.
In Fig.~\ref{Fig:TCF}(b), we observe an oscillatory behavior.

\section{Overdamped Langevin system with odd elastic tensor}
\label{Sec:Over}

\subsection{$N$ degrees of freedom}

As shown in Fig.~\ref{model}(b), we consider a deformable object such as an enzyme in a passive viscous fluid.
We investigate its active dynamics induced by the energy injection as a result of a chemical reaction.
We further assume that the odd elastic tensor can describe such a non-equilibrium process as argued in 
Refs.~\cite{Yasuda22, Yasuda21b}.
To discuss the structural fluctuation of the object driven by thermal motions of the surrounding passive fluid, 
we employ an overdamped Langevin system with an odd elastic tensor.

The Langevin equation for $N$ position variables $x_\alpha(t)$ can be written as~\cite{KuboBook,DoiBook,Weiss03,Weiss07}
\begin{align}
\dot x_\alpha=-M_{\alpha\beta}K_{\beta\gamma}x_\gamma+G_{\alpha\beta}\xi_\beta(t),
\label{OverLanEq}
\end{align}
where $M_{\alpha\beta}$ is the mobility tensor that is symmetric, 
$M_{\alpha\beta}=M_{\beta\alpha}$, due to Onsager's reciprocal relations for passive fluids.
Moreover, $M_{\alpha\beta}$ is positive definite according to the second law of thermodynamics~\cite{DoiBook}.
In general, the mobility tensor $M_{\alpha\beta}$ is the inverse of the resistance tensor $\Lambda_{\alpha\beta}$ 
introduced in the previous section, and $M_{\alpha\beta}$ can also be non-symmetric for active chiral fluids.
However, we do not consider such a general case because we intend to attribute the origin of the 
non-equilibrium effect to the chemical reaction and not to the activity in the surrounding fluid.
As in the previous section, $\xi_\beta(t)$ is Gaussian white noise that satisfies Eq.~(\ref{WGN}). 
The tensor $G_{\alpha\beta}$ represents the noise strength that is further related to the diffusion 
tensor by $D_{\alpha\beta}=G_{\alpha\gamma}G_{\beta\gamma}/2$ that is symmetric by 
definition.
In our work, we consider stochastic processes driven by thermal fluctuations
and assume the relation $D_{\alpha\beta}=k_{\rm B}TM_{\alpha\beta}$~\cite{KuboBook,DoiBook}.

In Eq.~(\ref{OverLanEq}), $K_{\alpha\beta}$ is the elastic constant tensor.
For passive systems, $K_{\alpha\beta}$ should be symmetric because elastic forces 
are conservative~\cite{LandauBook}. 
For active systems with non-conservative interactions, however, $K_{\alpha\beta}$ can have an anti-symmetric 
part that corresponds to odd elasticity~\cite{Scheibner20,Yasuda22,Yasuda21b}.
Hence $K_{\alpha\beta}$ can generally be written as 
\begin{align}
K_{\alpha\beta}=K_{\alpha\beta}^{\rm e}+K_{\alpha\beta}^{\rm o},
\label{EC-o}
\end{align}
where the symmetric (even) part and the anti-symmetric (odd) part satisfy 
$K_{\alpha\beta}^{\rm e}=K_{\beta\alpha}^{\rm e}$ 
and  $K_{\alpha\beta}^{\rm o}=-K_{\beta\alpha}^{\rm o}$, respectively,
similar to the resistance tensor.

For $N$-dimensional overdamped equations, we obtain the short-time behavior of $\phi_{\alpha\beta}(t)=
\phi_{\alpha\beta}^\mathrm S(t) + \phi_{\alpha\beta}^\mathrm A(t)$ as 
(see Eqs.~(\ref{TCF-ST-e}) and (\ref{TCF-ST-o}) in Appendix~\ref{App:A})
\begin{align}
\phi_{\alpha\beta}^\mathrm S(t) & \approx\bar\phi_{\alpha\beta}-k_\mathrm BTM_{\alpha\beta}|t|, 
\\
\phi_{\alpha\beta}^\mathrm A(t) & \approx-\frac{1}{2}[M_{\alpha\gamma}K_{\gamma\delta}\bar\phi_{\delta\beta}-M_{\beta\gamma}K_{\gamma\delta}\bar\phi_{\delta\alpha}]t=\bar\chi_{\alpha\beta}t,
\label{Gen-ST-TCF-phi}
\end{align}
where $\bar\chi_{\alpha\beta}=\langle v_\alpha x_\beta \rangle$ and 
notice the relation $\chi_{\alpha\beta}(t)=\dot\phi_{\alpha\beta}(t)$.
In the above, we have assumed that $|\overline{MK}_i||t|\ll 1$ is satisfied for all the eigenvalues
$\overline{MK}_i$ of the matrix $M_{\alpha\gamma}K_{\gamma\beta}$.
The equal-time-correlation function $\bar\phi_{\alpha\beta}=\langle x_\alpha x_\beta \rangle$ obeys the Lyapunov equation~\cite{Weiss03,Weiss07}: 
\begin{align}
M_{\alpha\gamma}K_{\gamma\delta}\bar\phi_{\delta\beta}+M_{\beta\gamma}K_{\gamma\delta}\bar\phi_{\delta\alpha}=2D_{\alpha\beta}.
\end{align}
We eliminate $\bar\phi_{\alpha\beta}$ from this Lyapunov equation by using $\bar\chi_{\alpha\beta}$ in Eq.~(\ref{Gen-ST-TCF-phi}), and obtain
\begin{align}
M_{\alpha\gamma}K_{\gamma\delta}\bar \chi_{\delta\beta}+\bar \chi_{\alpha\gamma} K_{\delta\gamma}M_{\delta\beta}=-2k_\mathrm BTM_{\alpha\gamma}K_{\gamma\delta}^\mathrm oM_{\delta\beta}.
\label{Gen-ST-TCF-phi-R}
\end{align}
With this equation, we can prove that $\bar\chi_{\alpha\beta}\ne0$ and hence $\phi_{\alpha\beta}^\mathrm A\ne0$
when $K_{\alpha\beta}^\mathrm o\ne0$.
Therefore the time-reversal symmetry discussed in Eq.~(\ref{TRSpsi}) is broken due to the presence of 
$K_{\alpha\beta}^\mathrm o$.

\subsection{Two degrees of freedom}

Alternatively, we discuss the time-correlation functions when $N=2$.
We assume that the elastic tensor is given by the following form:
\begin{align}
K_{\alpha\beta}=k^\mathrm e \delta_{\alpha\beta}+k^\mathrm o\epsilon_{\alpha\beta}.
\label{SymK}
\end{align} 
In the following, we further assume that the mobility tensor takes the form 
$M_{\alpha\beta}=\mu\delta_{\alpha\beta}$.
A more general situation is discussed in Appendix~\ref{App:C}.

The position-position correlation function 
$\phi_{\alpha\beta}(t)=\langle x_\alpha(t)x_\beta(0)\rangle = 
\phi_{\alpha\beta}^\mathrm S(t) + \phi_{\alpha\beta}^\mathrm A(t)$
is given by 
\begin{align}
\phi_{\alpha\beta}^\mathrm S(t) & =
\frac{k_\mathrm BT}{k^\mathrm e}e^{-\mu k^\mathrm e |t|} \cos(\mu k^\mathrm o t) \delta_{\alpha\beta},
\label{Over-phi-Sym0}
\\
\phi_{\alpha\beta}^\mathrm A(t) & = -
\frac{k_\mathrm BT}{k^\mathrm e}e^{-\mu k^\mathrm e |t|} \sin(\mu k^\mathrm o t) \epsilon_{\alpha\beta}.
\label{Over-phi-Asym}
\end{align}
Then the equal-time-correlation function becomes 
\begin{align}
\bar \phi_{\alpha\beta}&=\frac{k_\mathrm BT}{k^\mathrm e}\delta_{\alpha\beta}
\label{OverSTCF},
\end{align}
which is independent of $k^\mathrm o$.
As shown in Appendix~\ref{App:C}, however, $\bar \phi_{\alpha\beta}$ can depend on $k^\mathrm o$ in 
a more general situation.
The behavior of $\phi_{\alpha\beta}(t)$ is the same as that of $\psi_{\alpha\beta}(t)$ in Fig.~\ref{Fig:TCF}
as long as the proper scaling is made.

In the short-time limit, i.e., $|\mu k^\mathrm e t|\ll1$ and $|\mu k^\mathrm o t|\ll1$, 
Eqs.~(\ref{Over-phi-Sym0}) and (\ref{Over-phi-Asym}) become
\begin{align}
\phi_{\alpha\beta}^\mathrm S(t)&\approx\bar \phi_{\alpha\beta}-k_\mathrm BT \mu |t| \delta_{\alpha\beta},
\label{OverTCF-e-ST}
\\
\phi_{\alpha\beta}^\mathrm A(t) & \approx- \frac{k_\mathrm BT k^\mathrm o}{k^\mathrm e} \mu  t\epsilon_{\alpha\beta}.
\label{OverTCF-o-ST}
\end{align}
The slope of the symmetric part is given by the transport coefficient $\mu$  and hence it is related to 
the diffusion coefficient according to the fluctuation dissipation theorem~\cite{DoiBook}.
On the other hand, the slope of the anti-symmetric part is characterized by the ratio $k^\mathrm o/k^\mathrm e$.

\begin{figure}[tb]
\begin{center}
\includegraphics[scale=0.5]{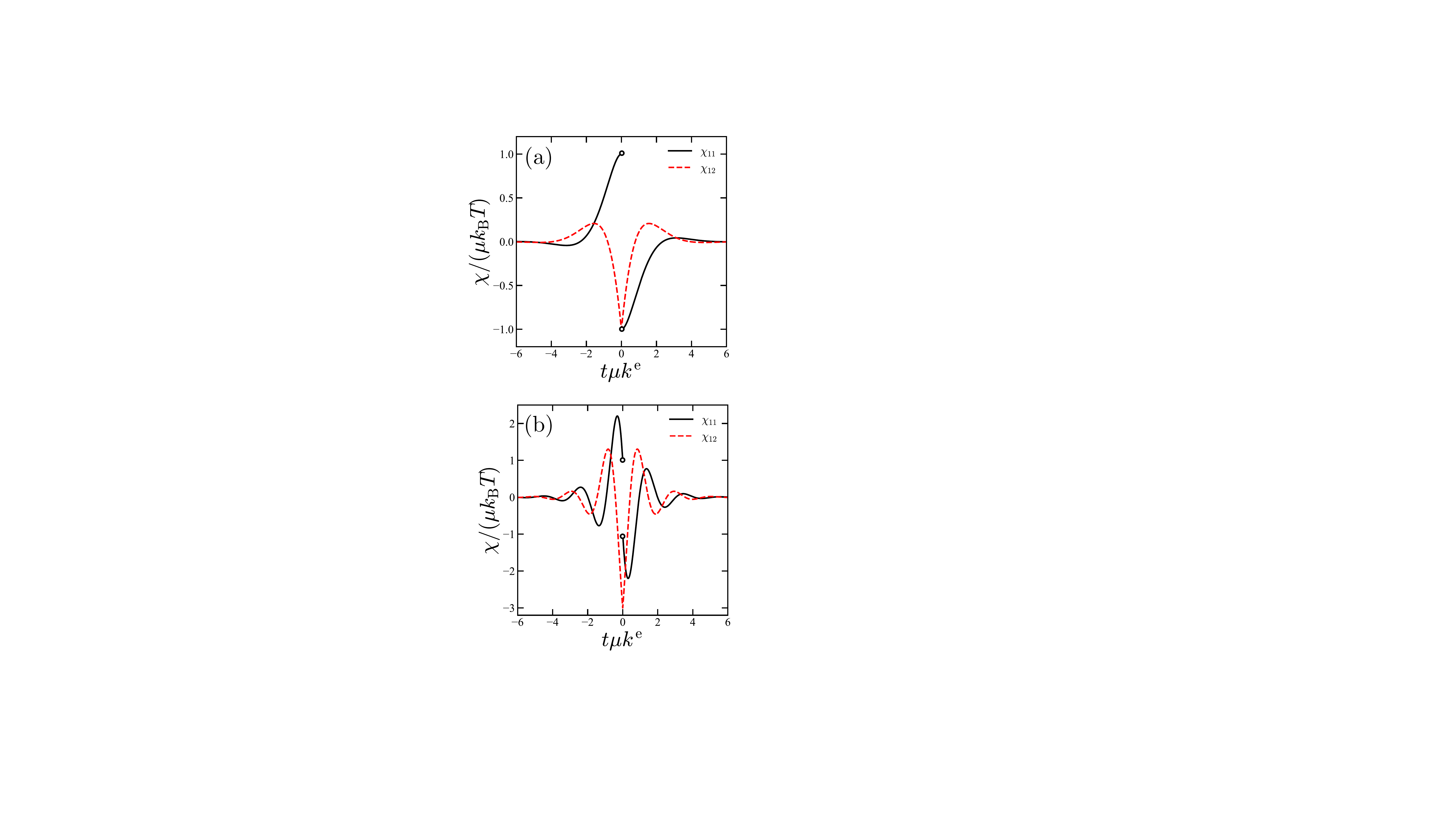}
\end{center}
\caption{
Plots of the scaled velocity-velocity correlation functions $\chi_{11}(t)$ (black solid line) and 
$\chi_{12}(t)$ (red dashed line) as a function of dimensionless time $t \mu k^\mathrm e$
when (a) $k^\mathrm o/k^\mathrm e=1$ and (b) $k^\mathrm o/k^\mathrm e=3$ 
for $N=2$ [see Eqs.~(\ref{Over-chi-Sym0}) and (\ref{Over-chi-Asym})].
In both cases, $\chi_{11}(t)$ is an odd function of time and discontinuous at $t=0$, while 
$\chi_{12}(t)$ is an even function of time.
Since $\chi_{12}(t)$ violates Eq.~(\ref{TRSchi}), time-reversal invariance is broken 
in the presence of odd elastic tensor.
In (b), we observe oscillatory behavior of the correlation functions. 
}
\label{Fig:TCF2}
\end{figure}

Under the same assumptions, we next discuss the position-velocity correlation function  
$\chi_{\alpha\beta}(t)=\langle v_\alpha(t) x_\beta(0)\rangle = 
\chi_{\alpha\beta}^\mathrm S(t) + \chi_{\alpha\beta}^\mathrm A(t)$, where
\begin{align}
\chi_{\alpha\beta}^\mathrm S(t) & = -k_\mathrm BT\mu e^{-\mu k^\mathrm e |t|} 
\nonumber \\
& \times \left[\mathrm{Sgn}\,(t)\cos(\mu k^\mathrm o t) +\frac{k^\mathrm o}{k^\mathrm e}\sin(\mu k^\mathrm o t) \right] \delta_{\alpha\beta}, 
\label{Over-chi-Sym0}
\\
\chi_{\alpha\beta}^\mathrm A(t) & =
k_\mathrm BT\mu e^{-\mu k^\mathrm e |t|} 
\nonumber \\
& \times \left[\sin(\mu k^\mathrm o |t|)- \frac{k^\mathrm o}{k^\mathrm e}\cos(\mu k^\mathrm o t)\right] \epsilon_{\alpha\beta}.
\label{Over-chi-Asym}
\end{align}
In the above, the function $\mathrm {Sgn}\,(t)=t/|t|$ takes either $1$ or $-1$ depending on its sign.
In accordance with Eq.~(\ref{TTSchi-EO}), $\chi_{\alpha\beta}^\mathrm S(t)$ is an odd function of time, 
while $\chi_{\alpha\beta}^\mathrm A(t)$ is an even function.
On the other hand, the presence of $\chi_{\alpha\beta}^\mathrm A$ indicates the broken time-reversal 
symmetry due to odd elasticity.

From Eq.~(\ref{Over-chi-Asym}), the equal-time-correlation function can be obtained as 
\begin{align}
&\bar \chi_{\alpha\beta}=-\frac{k_\mathrm BT k^\mathrm o}{k^\mathrm e} \mu  \epsilon_{\alpha\beta},
\label{Same-chi}
\end{align}
where we have used $\mathrm{Sgn}\,(0)=0$.
Since $\bar \chi_{\alpha\beta}^{\rm eq}=0$ should hold when time-reversal invariance is satisfied, the presence 
of $\bar \chi_{\alpha\beta}$ indicates that time-reversal symmetry is broken in the presence of odd elasticity.
In the short-time limit, as we considered in Eqs.~(\ref{OverTCF-e-ST}) and (\ref{OverTCF-o-ST}), 
Eqs.~(\ref{Over-chi-Sym0}) and (\ref{Over-chi-Asym}) become
\begin{align}
\chi_{\alpha\beta}^\mathrm S(t)&\approx -k_\mathrm BT\mu\left[\mathrm{Sgn}\,(t)-\mu \frac{(k^\mathrm e)^2-(k^\mathrm o)^2}{k^\mathrm e} t \right]\delta_{\alpha\beta}, 
\label{OverTCFchi-e-ST}
\\
\chi_{\alpha\beta}^\mathrm A(t) & \approx \bar\chi_{\alpha\beta}+2k_\mathrm BT\mu^2 k^\mathrm o |t|\epsilon_{\alpha\beta}.
\label{OverTCFchi-o-ST}
\end{align}

In Fig.~\ref{Fig:TCF2}, we plot $\chi_{11}(t)$ (black solid line) and $\chi_{12}(t)$ (red dashed line) as a 
function of dimensionless time $\mu k^\mathrm e t$.
The other parameter is $k^\mathrm o/k^\mathrm e=1$ in Fig.~\ref{Fig:TCF2}(a) and $k^\mathrm o/k^\mathrm e=3$
in Fig.~\ref{Fig:TCF2}(b).
In both cases, $\chi_{11}(t)$ is an odd function, while $\chi_{12}(t)$ is an even function.
In Fig.~\ref{Fig:TCF2}(b), we also observe an oscillatory behavior.

\begin{figure}[tb]
\begin{center}
\includegraphics[scale=0.55]{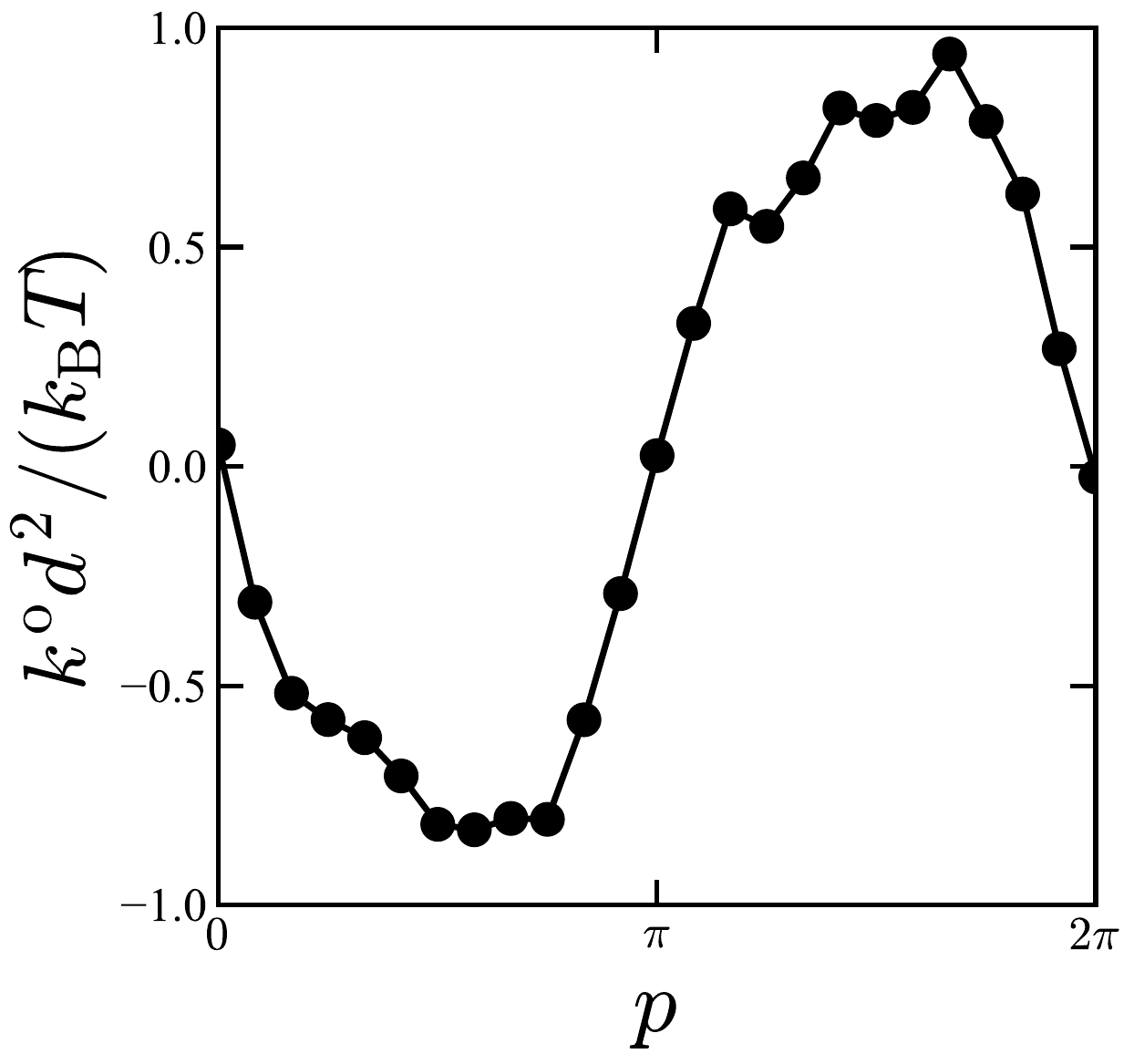}
\end{center}
\caption{
Plot of the odd elasticity $k^\mathrm o$ estimated from the numerical simulations of the 
enzyme model in Ref.~\cite{Yasuda21a} as a function of the phase difference $p$.
The chosen parameters are $\nu/(k_\mathrm BT)=16$, $h/(k_\mathrm BT)=20$, $cd^2/(k_\mathrm BT)=20$, 
and $\mu_s/(\mu_\theta d^2)=1$.
}
\label{Fig:Odd-elasticity}
\end{figure}

\section{Odd elasticity of an enzyme system with chemical reaction}
\label{Sec:enzyme}

In this section, we discuss the application of our results to structural fluctuations in a model enzyme 
system introduced in Refs.~\cite{Adeleke19,Hosaka20,Yasuda21a,Canalejo21} and also shown in Fig.~\ref{model}(b).
An enzyme changes its shape during a catalytic chemical reaction and receives chemical energy from 
a substrate molecule.
This process is quite complicated and many degrees of freedom such as the positions of 
all the atoms are involved. 
To tackle such a problem, we first coarse grain the system to obtain the dynamical equations 
with minimum degrees of freedom such as in Eq.~(\ref{OverLanEq}). 
Then we assume that the energy injection from the substrate molecule is effectively described by the 
odd part of the elastic tensor.
As an example of the application of our analytical results, we use the time correlation functions to 
obtain the effective odd elasticity of the enzyme model proposed by us~\cite{Yasuda21a}.

In our model, we consider the dynamics of the extent of catalytic reaction 
$\theta(t)$ and the structure of an enzyme characterized by $s_1(t)$ and $s_2(t)$.
The free energy describing a chemical reaction is given by
$g_{\mathrm{r}}(\theta)=-h\cos\theta-\nu\theta$, where $h$ is the 
energy barrier in the chemical reaction and $\nu$ is the chemical potential difference.
We also introduce the following mechano-chemical coupling energy
\begin{align}
g_\mathrm{c}(\theta,\{s_i\})=\frac{c}{2}\left([s_1-d\sin\theta]^2+[s_2-d\sin(\theta+p)]^2\right),
\label{Gs}
\end{align}
where $c$ is the coupling strength, $d$ is the amplitude of the structure change, and $p$ is the 
phase difference relative to the reaction phase.
The total free energy is given by 
$g_{\mathrm{t}}(\theta,\{s_i\})=g_\mathrm{r}(\theta)+g_\mathrm{c}(\theta,\{s_i\})$.
The Onsager's phenomenological equations
\begin{align}
\dot{\theta}& =-\mu_{\theta}\partial_\theta g_{\mathrm{t}}+\sqrt{2\mu_\theta}\xi, 
\\
\dot{s_i}& =-\mu_s\partial_{s_i} g_{\mathrm{t}}+ \sqrt{2\mu_s}\xi_i
\end{align}
determine the time evolution of each variable.
Here, $\xi$ and $\xi_i$ represent thermal fluctuations that satisfy Eq.~(\ref{WGN}).
A more detailed explanation of our enzyme model is provided in Ref.~\cite{Yasuda21a}.

We have performed numerical simulations of the above Langevin dynamics and calculated the structural 
time correlation functions.
We find that the time-reversal symmetry in Eq.~(\ref{TRSphi}) is broken in the correlation function 
$\phi_{12}(t)\sim\langle (s_1(t)-\langle s_1\rangle)(s_2(0)-\langle s_2\rangle)\rangle$ when $\nu\ne0$.
Comparing the short time behavior of the simulation result with Eq.~(\ref{app:OverTCF-gen-ST}), which is a 
generalized expression of Eq.~(\ref{OverTCF-o-ST}), we have estimated the effective odd elasticity of the 
enzyme system.
In Fig.~\ref{Fig:Odd-elasticity}, we plot the odd elasticity $k^\mathrm o$ as a function of $p$ and 
find a periodic dependence that can be approximately described by $k^\mathrm o\sim-\sin p$.
In our model, the phase difference $p$ between the structural variables $s_1$ and $s_2$ 
is introduced  to account for the non-reciprocal deformation of an enzyme molecule.
Such a non-reciprocality can be quantitatively characterized by the area enclosed by a trajectory in a space 
spanned by $s_1$ and $s_2$~\cite{Yasuda21a}. 
The relation $k^\mathrm o\sim-\sin p$ indicates that the effective odd elasticity of an enzyme can be 
obtained only by measuring the structural dynamics without assuming any detailed dynamics of the internal 
variable such as $\theta$.

A more detailed analysis of the simulation results will be presented in a separate publication. 
We emphasize here that such an analysis suggests a new possibility to understand non-equilibrium 
dynamics of active matter.

\section{Summary and discussion}
\label{Sec:Dis}

In this paper, we have investigated the statistical properties of fluctuations in active systems that are governed by 
non-symmetric responses. 
We first summarized the symmetry properties of the time-correlation matrices due to time-translational 
and time-reversal invariances. 
The anti-symmetric parts of the time-correlation functions can exist in non-equilibrium situations.
We investigated an underdamped Langevin system with a non-symmetric resistance tensor and obtained the time 
correlation matrices.  
We showed that time-reversal symmetry is violated in the presence of the odd part of the resistance tensor.
For a system with two degrees of freedom, we calculated the analytical expressions of the time-correlation 
functions.

Next, we discussed an overdamped Langevin system with a non-symmetric elastic tensor and obtained the 
corresponding time-correlation functions.
We also showed that time-reversal symmetry of the correlation functions is violated in the presence of 
odd elasticity. 
The initial slope of the time-correlation functions represent the transport coefficient and the odd elasticity 
for the symmetric and the asymmetric parts, respectively.
In particular, the position-velocity correlation function typically reflects the broken time-reversal 
symmetry and is proportional to the odd elasticity.

Let us give some numerical estimates of the physical quantities used in the present work.
We consider the case when the concept of odd elasticity is applied to the structural changes of 
enzymes and motor proteins. 
The domain size of a protein is $a \approx 10^{-8}$~m and the viscosity of water is 
$\eta \approx 10^{-3}$\,Pa$\cdot$s.
Hence the transport coefficient becomes is 
$\mu = 1/(6\pi\eta a) \approx 5\times10^9$\,m$^2$/(J$\cdot$s).
According to the experiments on a kinesin molecule~\cite{Ariga18}, the even elasticity can be roughly estimated 
as $k^\mathrm e \approx 1\times 10^{-4}$\,J/m$^2$.
Then the relaxation rate can be roughly estimated as $\mu k^\mathrm e \approx 5\times10^5$\,s$^{-1}$.
Next, we estimate the odd elastic constant from the activity of motor proteins.
The active force due to kinesin is estimated to be $f \approx 10^{-11}$\,N~\cite{Kojima97,Ariga18}.
By estimating the rough displacement to be $d \approx 10^{-8}$~m, 
the odd elastic constant can be estimated as $k^\mathrm o\sim f/d \approx 10^{-3}$\,J/m$^2$.
Then the ratio between the odd and even elastic constants can be typically 
$k^\mathrm o/k^\mathrm e \approx 10$.

In Secs.~\ref{Sec:Under} and \ref{Sec:Over}, we introduced the symmetric tensors $B_{\alpha\beta}$ and 
$D_{\alpha\beta}$ representing the noise strength.
For general active situations, the noise originates not only from thermal fluctuations but also from 
non-equilibrium fluctuations.
In active cases, $B_{\alpha\beta}$ and $D_{\alpha\beta}$ do not need to obey the fluctuation dissipation theorem
and they are general positive definite symmetric tensors.
Other generalization is to take into account the anti-symmetric parts of $B_{\alpha\beta}$ and $D_{\alpha\beta}$.
In this situation, however, the time-reversal symmetry of noise [see Eq.~(\ref{WGN})] can also be broken, and 
such a generalization is beyond the scope of the present work.

In Secs.~\ref{Sec:Under} and \ref{Sec:Over} we have independently investigated the systems 
with non-symmetric resistance tensor and non-symmetric elastic tensor.
When both of these properties exist simultaneously, the general Langevin equation can be written as 
\begin{align}
m\ddot x_\alpha=-\Lambda_{\alpha\beta}\dot x_\beta-K_{\alpha\beta}x_\beta+mF_{\alpha\beta}\xi_\beta(t),
\end{align}
where $\Lambda_{\alpha\beta}$ and $K_{\alpha\beta}$ are given by Eqs.~(\ref{DC-o}) and (\ref{EC-o}), 
respectively.
If the separation of time scales $m/\Lambda \ll  1/(\mu K)$ holds in a mesoscopic system, 
one can eliminate the inertia term and obtain the overdamped Langevin equation~\cite{DoiBook}. 
In a thermal equilibrium system, the velocity-velocity correlation function $\psi_{\alpha\beta}(t)$ becomes a  
delta function in the limit of $m/\Lambda \to 0$. 
In non-equilibrium systems with odd properties, however, the time-reversal symmetry of $\psi_{\alpha\beta}(t)$
is violated as in Eq.~(\ref{Gen-ST-TCF-psi}).
Hence the time-correlation functions of the noise terms should involve not only the delta function  
but also its time derivative as discussed in Ref.~\cite{Han21}.

Recently, the variational principle for active matter has been proposed~\cite{Wang21}.
Within such an extended variational principle, one can obtain the dynamical equation in Eq.~(\ref{OverLanEq}) 
by minimizing the Rayleighian $\mathcal{R}=(M^{-1})_{\alpha\beta} \dot x_\alpha\dot x_\beta/2+ \dot{A}
+\dot {W}$, where $(M^{-1})_{\alpha\beta}$ is the inverse matrix of $M_{\alpha\beta}$.
The first and second terms are the dissipation function and the time derivative of the free energy, respectively,
while $\dot W$ is the time derivative of the work generated by active forces. 
In our work, the even and odd elastic tensors can be included within the variational principle by choosing 
$\dot A=K_{\alpha\beta}^\mathrm e\dot x_\alpha x_\beta$ and 
$\dot W=K_{\alpha\beta}^\mathrm o\dot x_\alpha x_\beta$
to obtain Eq.~(\ref{OverLanEq}).
On the other hand, the odd resistance tensor cannot be described within the variational principle.

In the present work, we have discussed the linear Langevin equations with odd resistance tensor or odd elastic tensor. 
It should be noted, however, we can also discuss nonlinear effects by considering a state-dependent
resistance tensor $\Lambda_{\alpha\beta}^\mathrm e(\{x_\alpha\})$ or a state-dependent mobility tensor 
$\mu_{\alpha\beta}(\{x_\alpha\})$ as well as nonlinear conservative forces. 
For example, nonlinearity appears in the dynamics of a deformable object in the presence of hydrodynamic 
interactions~\cite{Yasuda21b,Ishimoto21}.
For the future, the study of nonlinear dynamics in the presence of odd properties is necessary.

In this paper, we have discussed the Langevin systems with either odd resistance tensor
or odd elastic tensor.
Owing to the mathematical analogy, the obtained results for the odd Langevin systems have various applications.
Examples of the Langevin system with odd resistance tensor are 
the Brownian particle in a chiral active fluid (see Sec.~\ref{Sec:Under}), and 
the Brownian particle under the Lorentz-forces~\cite{Sabass17}.
On the other hand, odd elastic tensor can exist such as in 
the structural dynamics of a catalytic enzyme (see Sec.~\ref{Sec:enzyme}),
the Brownian particle in shear flows~\cite{Holzer10}, 
and the stochastic behavior of the climate system~\cite{Weiss20}.

To further strengthen our idea of effective odd elasticity, the following approaches will be useful.
If we can generalize the projection operator formulation to non-equilibrium systems with chemical 
reactions, the coarse-grained Langevin equation such as Eq.~(\ref{OverLanEq}) can be obtained 
from the Hamilton dynamics by including all degrees of 
freedom~\cite{Kawasaki73,ZwanzigBook,Izvekov21} .
On the other hand, numerical simulations based on multi-particle collision dynamics can also be used 
to investigate the time-correlation functions of the enzyme structure~\cite{Echeverria11,Echeverria12}.

\section*{Data availability statements}

The data that support the findings of this study are available from the corresponding author upon reasonable request.

\begin{acknowledgments}

We thank M.\ Doi, H.\ Hayakawa,  J.\ B.\ Weiss, and A.\ Zaccone for useful comments. 
K.Y.\ acknowledges the support by a Grant-in-Aid for JSPS Fellows (Grant No.\ 21J00096) from the 
Japan Society for the Promotion of Science (JSPS).
K.Y.\ was supported by the Research Institute for Mathematical Sciences, an International 
Joint Usage/Research Center located in Kyoto University.
K.I.\ acknowledges the JSPS, KAKENHI for Transformative Research Areas A (Grant No.\ 21H05309) 
and the Japan Science and Technology Agency (JST), PRESTO Grant (No.\ JPMJPR1921).
S.K.\ acknowledges the supported by the startup fund of Wenzhou Institute, University of Chinese Academy 
of Sciences (No.\ WIUCASQD2021041). 
K.Y.\ and K.I.\ were supported by the Research Institute for Mathematical Sciences, an International 
Joint Usage/Research Center located in Kyoto University.
\end{acknowledgments}

\appendix
\begin{widetext}
\section{General analysis of the linear Langevin equation}
\label{App:A}

In this Appendix, we provide a general method to solve the linear Langevin equation with 
$N$ degrees of freedom given by~\cite{Weiss03,Weiss07}
\begin{align}
\dot X_\alpha=A_{\alpha\beta}X_\beta+H_{\alpha\beta}\xi_\beta(t),
\label{GenLinLanEq}
\end{align}
where $X_\alpha$ represents any set of variables such as $x_\alpha$ or $v_\alpha$,
and $A_{\alpha\beta}$ is the $N\times N$ tensor characterizing the decay rate of the linear system. 
For the stability of the system, the real part of the all eigenvalues of $A_{\alpha\beta}$ must 
be negative. 
The amplitude of the noise is given by a $N\times N$ tensor $H_{\alpha\beta}$ whose diagonal 
and off-diagonal components correspond to auto-correlations and cross-correlations of the noise, 
respectively. 
Moreover, $\xi_\beta(t)$ is Gaussian white noise that satisfies Eq.~(\ref{WGN}).
The above Langevin equation can be formally solved as 
\begin{align}
X_\alpha(t)=(e^{At})_{\alpha\beta}\int_{-\infty}^tds\,(e^{-As})_{\beta\gamma}H_{\gamma\delta}\xi_\delta(s),
\end{align}
where we have used the matrix exponential $(e^{At})_{\alpha\beta}$.
We can immediately confirm $\langle X_\alpha(t)\rangle=0$.

Using the above solution, we can calculate the time-correlation functions 
$\Phi_{\alpha\beta}(t)=\langle X_\alpha(t)X_\beta(0)\rangle$ as  
\begin{align}
& \Phi_{\alpha\beta}(t) =\left\{ 
\begin{array}{ll}
(e^{At})_{\alpha\gamma}\bar\Phi_{\gamma\beta} & (t\ge0) \\
\bar\Phi_{\alpha\gamma}(e^{-At})_{\beta\gamma} & (t<0)
\end{array}
\right.,\\
& \bar\Phi_{\alpha\beta} = 2\int_{0}^\infty ds\,(e^{As})_{\alpha\gamma}C_{\gamma\delta}(e^{As})_{\beta\delta},
\label{Same-time-corr}
\end{align}
where $\bar\Phi_{\alpha\beta}=\Phi_{\alpha\beta}(0)$ are the equal-time-correlation functions
and $C_{\alpha\beta}=H_{\alpha\gamma}H_{\beta\gamma}/2$.
Notice that $\bar\Phi_{\alpha\beta}$ obeys the following Lyapunov equation: 
\begin{align}
A_{\alpha\gamma}\bar\Phi_{\gamma\beta}+A_{\beta\gamma}\bar\Phi_{\gamma\alpha}+2C_{\alpha\beta}=0.
\label{Lyapunov-eq}
\end{align}
The integral in Eq.~(\ref{Same-time-corr}) can be calculated when $A_{\alpha\beta}$ is diagonalized by 
a matrix $P_{\alpha\beta}$. 
Then, $\bar\Phi_{\alpha\beta}$ is obtained by using the eigenvalues $\zeta_i$ of the 
matrix $A_{\alpha\beta}$ as 
\begin{align}
\bar\Phi_{\alpha\beta}& =-2P_{\alpha\gamma}P_{\beta\delta}O_{\gamma\delta},
\\
O_{\gamma\delta}&=(P^{-1})_{\gamma l}C_{lm}(P^{-1})_{\delta m}\circ \frac{1}{\zeta_\gamma+\zeta_\delta},
\end{align}
where $\circ$ stands for the Hadamard product (element-wise product).
We note that $A_{\alpha\beta}$ cannot always be diagonalized when $A_{\alpha\beta}$ is a non-symmetric matrix.

The time-correlation functions can be decomposed into the symmetric and anti-symmetric parts as
\begin{align}
\Phi_{\alpha\beta}(t)&=\Phi_{\alpha\beta}^\mathrm S(t)+\Phi_{\alpha\beta}^\mathrm A(t),
\label{TCF}\\
\Phi_{\alpha\beta}^\mathrm S(t)&=\frac{1}{2}\left[(e^{A|t|})_{\alpha\gamma}\bar\Phi_{\gamma\beta}+(e^{A|t|})_{\beta\gamma}\bar\Phi_{\gamma\alpha}\right],
\label{TCF-e}\\
\Phi_{\alpha\beta}^\mathrm A(t)&=\frac{1}{2}\mathrm {Sgn}(t)\left[(e^{A|t|})_{\alpha\gamma}\bar\Phi_{\gamma\beta}-(e^{A|t|})_{\beta\gamma}\bar\Phi_{\gamma\alpha}\right],
\label{TCF-o}
\end{align}
for both negative and positive $t$.
We confirm here $\Phi_{\alpha\beta}^\mathrm S(t)=\Phi_{\alpha\beta}^\mathrm S(-t)$ and 
$\Phi_{\alpha\beta}^\mathrm A(t)=-\Phi_{\alpha\beta}^\mathrm A(-t)$~\cite{KuboBook}.
Moreover, we have $\bar\Phi_{\alpha\beta}=\bar\Phi_{\beta\alpha}$. 
When $\bar\Phi_{\alpha\beta}\propto \delta_{\alpha\beta}$ and $A_{\alpha\beta}\ne A_{\beta\alpha}$, 
we can easily confirm $\Phi_{\alpha\beta}^\mathrm A(t)\ne0$, which indicates the time-reversal symmetry breaking.
The short-time asymptotic expressions of Eqs.~(\ref{TCF-e}) and (\ref{TCF-o}) become 
\begin{align}
\Phi_{\alpha\beta}^\mathrm S(t)&\approx \bar\Phi_{\alpha\beta}+\frac{1}{2}\left[A_{\alpha\gamma}\bar\Phi_{\gamma\beta}+A_{\beta\gamma}\bar\Phi_{\gamma\alpha}\right]|t|=\bar\Phi_{\alpha\beta}-C_{\alpha\beta}|t|,
\label{TCF-ST-e}\\
\Phi_{\alpha\beta}^\mathrm A(t)&\approx \frac{1}{2}[A_{\alpha\gamma}\bar\Phi_{\gamma\beta}-A_{\beta\gamma}\bar\Phi_{\gamma\alpha}]t.
\label{TCF-ST-o}
\end{align}

For two degrees of freedom ($N=2$), one can solve the Langevin equation analytically and 
$\bar \Phi_{\alpha\beta}$ is given by
\begin{align}
\bar \Phi_{\alpha\beta}&=-\frac{C_{\alpha\beta}}{\mathrm {tr}[A]}-\frac{\det[A]}{\mathrm {tr}[A]}(A^{-1})_{\alpha\gamma}C_{\gamma\delta}(A^{-1})_{\beta\delta}.
\label{2DSTCF}
\end{align}
The time-dependence of the correlation functions can be obtained as
 \begin{align}
(e^{A t})_{\alpha\beta}=e^{\gamma t}
\left[ \cosh(\omega t)\delta_{\alpha\beta}
+\frac{A_{\alpha\beta}-\det[A] (A^{-1})_{\alpha\beta}}{2\omega}\sinh(\omega t)\right],
\label{2DMatExp}
\end{align}
where we have introduced the relaxation rate $\gamma=\mathrm {tr}[A]/2$ and the 
frequency $\omega=\sqrt{\gamma^2-\det[A]}$.

\section{Correlation functions in underdamped Langevin systems}
\label{App:B}

In this Appendix, we give the general expressions of the time-correlation functions for an 
underdamped system when $N=2$.
Here $\Lambda_{\alpha\beta}^\mathrm e$ is a symmetric and positive definite $2\times2$ matrix,
while the odd part of the resistance tensor is given by 
$\Lambda_{\alpha\beta}^\mathrm o=\lambda^\mathrm o\epsilon_{\alpha\beta}$.
Comparing Eqs.~(\ref{UnderLanEq}) and (\ref{GenLinLanEq}), we obtain 
$A_{\alpha\beta}=-\Lambda_{\alpha\beta}/m$.
From Eqs.~(\ref{TCF-e}) and (\ref{TCF-o}), we then have 
\begin{align}
\psi_{\alpha\beta}^\mathrm S(t)&=\frac{k_\mathrm BT}{2m}\left[(e^{-\Lambda|t|/m})_{\alpha\beta}+(e^{-\Lambda|t|/m})_{\beta\alpha}\right],\\
\psi_{\alpha\beta}^\mathrm A(t)&=\mathrm {Sgn}\,(t)\frac{k_\mathrm BT}{2m}\left[(e^{-\Lambda|t|/m})_{\alpha\beta}-(e^{-\Lambda|t|/m})_{\beta\alpha}\right],
\end{align}
where we have used the equipartition theorem in Eq.~(\ref{equipartition}).

Furthermore, using Eq.~(\ref{2DMatExp}) and the relation
$(\Lambda^{-1})_{\alpha\beta}=[\det[\Lambda^\mathrm e]((\Lambda^\mathrm e)^{-1})_{\alpha\beta}-\lambda^\mathrm o\epsilon_{\alpha\beta}]/\det[\Lambda]$ for a $2\times2$ matrix, we obtain 
\begin{align}
\psi_{\alpha\beta}^\mathrm S(t)& =\frac{k_\mathrm BT}{m}e^{- \gamma|t|}
\left[\cosh(\omega t)\delta_{\alpha\beta}
-\frac{\sinh(\omega|t|)}{2\omega m}
\left[\Lambda_{\alpha\beta}^\mathrm e-\det[\Lambda^\mathrm e]((\Lambda^\mathrm e)^{-1})_{\alpha\beta}\right]\right],
\label{2DTCF-psi-e}\\
\psi_{\alpha\beta}^\mathrm A(t)& =-\frac{k_\mathrm BT}{m^2\omega} \lambda^\mathrm o e^{- \gamma|t|}
\sinh(\omega t)\epsilon_{\alpha\beta}.
\label{2DTCF-psi-o}
\end{align}
In the above, we have defined the relaxation rate $\gamma=\mathrm{tr} [\Lambda^\mathrm e]/(2m)$ 
and the frequency $\omega=\sqrt{\gamma^2m^2-\det[\Lambda^\mathrm e]-(\lambda^\mathrm o)^2}/m$.

The exceptional point is given by the condition $\omega=0$ and we obtain 
$\lambda^{\rm o}_{\rm ep}=\pm\sqrt{\gamma^2m^2-\det[\Lambda^\mathrm e]}$~\cite{Fruchart21}.
The frequency $\omega$ is a real number when $(\lambda^\mathrm o)^2<(\lambda^{\rm o}_{\rm ep})^2$, 
while it is an imaginary number when $(\lambda^\mathrm o)^2>(\lambda^{\rm o}_{\rm ep})^2$ for 
which the time-correlation function can oscillate. 
To see an oscillating behavior, however, we further need a condition  
$\omega^2<-\gamma^2$, as shown in Fig.~\ref{Fig:TCF}(b).

\section{Correlation functions in overdamped Langevin systems}
\label{App:C}

In this Appendix, we give the general expressions of the time-correlation functions for an overdamped 
system when $N=2$.
Here both $M_{\alpha\beta}$ and $K_{\alpha\beta}^\mathrm e$ are symmetric and positive definite 
$2\times2$ matrices, while the odd part of the elastic tensor is given by $K_{\alpha\beta}^\mathrm o=k^\mathrm o\epsilon_{\alpha\beta}$. 
Comparing Eqs.~(\ref{OverLanEq}) and (\ref{GenLinLanEq}), we obtain 
$A_{\alpha\beta}=-M_{\alpha\gamma}K_{\gamma\beta}$ and $C_{\alpha\beta}=D_{\alpha\beta}=k_\mathrm BTM_{\alpha\beta}$.
From Eq.~(\ref{2DSTCF}), $\bar \phi_{\alpha\beta}$ becomes
\begin{align}
\bar \phi_{\alpha\beta}&=\frac{k_\mathrm BTM_{\alpha\beta}}{\mathrm {tr}[M K^\mathrm e]}+k_\mathrm BT\frac{\det[M] [\det [K^\mathrm e]+(k^\mathrm o)^2]}{\mathrm {tr}[M K^\mathrm e]}(K^{-1})_{\alpha\gamma}(M^{-1})_{\gamma\delta} (K^{-1})_{\beta\delta}.
\end{align}

Furthermore, using the relation $(K^{-1})_{\alpha\beta}=[\det[K^\mathrm e]((K^\mathrm e)^{-1})_{\alpha\beta}-k^\mathrm o\epsilon_{\alpha\beta}]/[\det[K^\mathrm e](1+\nu^2)]$ with $\nu^2=(k^\mathrm o)^2/\det[K^\mathrm e]$, we obtain the following expression
\begin{align}
\bar \phi_{\alpha\beta}&=\frac{k_\mathrm BT}{1+\nu^2}\left[((K^\mathrm e)^{-1})_{\alpha\beta}+\frac{2\nu^2}{\mathrm{tr}[M K^\mathrm e]}M_{\alpha\beta}-\frac{k^\mathrm o\det[M]}{\mathrm{tr}[M K^\mathrm e]}\left[\epsilon_{\alpha\gamma}(M^{-1})_{\gamma\delta} ((K^\mathrm e)^{-1})_{\delta\beta}+\epsilon_{\beta\gamma}(M^{-1})_{\gamma\delta} ((K^\mathrm e)^{-1})_{\delta\alpha}\right]\right].
\label{app:ETCF-ODLE}
\end{align}
In the above, we have used the identities $M_{\alpha\beta}+\det[M](M^{-1})_{\alpha\beta}=\mathrm{tr}[M]\delta_{\alpha\beta}$ and 
$\det[M]\epsilon_{\alpha\gamma}(M^{-1})_{\gamma\delta}\epsilon_{\beta\delta}=M_{\beta\alpha}$ for a $2\times 2$ matrix.
When $k^\mathrm o=0$ and hence $\nu=0$, the above expression reduces to 
$\bar \phi_{\alpha\beta}=k_\mathrm BT((K^\mathrm e)^{-1})_{\alpha\beta}$ corresponding to the thermal 
equilibrium case.

The time-dependence of the correlation functions is calculated by using Eq.~(\ref{2DMatExp}) 
\begin{align}
(e^{-M K|t|})_{\alpha\beta}&=
e^{- \gamma|t|}\left[\cosh(\omega t)\delta_{\alpha\beta}-\frac{M_{\alpha\delta} K_{\delta\beta}-\det[M]\det[K] (K^{-1})_{\alpha\delta}(M^{-1})_{\delta\beta}}{2\omega}\sinh(\omega |t|)\right],
\label{app:TD}
\end{align}
where we have introduced the relaxation rate $\gamma=\mathrm{tr}[M K^\mathrm e]/2$ and the 
frequency $\omega=\sqrt{\gamma^2-\det[M]\det[K^\mathrm e](1+\nu^2)}$.

The symmetric and anti-symmetric parts of the correlation matrix can be obtained from Eqs.~(\ref{TCF-e}) 
and (\ref{TCF-o}), respectively, as 
\begin{align}
& \phi_{\alpha\beta}^\mathrm S(t) =e^{- \gamma|t|}\cosh(\omega t)\bar \phi_{\alpha\beta}
\nonumber\\
& -\frac{k_\mathrm BTe^{-\gamma |t|}\sinh(\omega |t|)}{2\omega (1+\nu^2)\mathrm{tr}[M K^\mathrm e]}\left[M_{\alpha\gamma}K_{\gamma\delta}^\mathrm e M_{\delta\beta} 
-(\det[M])^2 (\det[K^\mathrm e])^2((K^\mathrm e)^{-1})_{\alpha\gamma}(M^{-1})_{\gamma\delta}((K^\mathrm e)^{-1})_{\delta\ell}(M^{-1})_{\ell m}((K^\mathrm e)^{-1})_{m \beta} \right.
\nonumber\\
& \left.+k^\mathrm o(\det[M])^2 \det[K^\mathrm e]\left[\epsilon_{\alpha\gamma}(M^{-1})_{\gamma\delta}((K^\mathrm e)^{-1})_{\delta \ell}(M^{-1})_{\ell m}((K^\mathrm e)^{-1})_{m\beta}+\epsilon_{\beta\gamma}(M^{-1})_{\gamma\delta}((K^\mathrm e)^{-1})_{\delta \ell}(M^{-1})_{\ell m}((K^\mathrm e)^{-1})_{m\alpha}\right]\right],
\label{app:Phi-e}
\end{align}
\begin{align}
\phi_{\alpha\beta}^\mathrm A(t)=-\frac{2k_\mathrm BTe^{- \gamma|t|}k^\mathrm o\sinh(\omega t)}
{\omega\,\mathrm{tr}[M K^\mathrm e]}\det[M] \epsilon_{\alpha\beta}.
\label{app:Phi-o}
\end{align}

In the short-time limit, Eqs.~(\ref{app:Phi-e}) and (\ref{app:Phi-o}) asymptotically become 
\begin{align}
\phi_{\alpha\beta}^\mathrm S(t)
\approx\bar \phi_{\alpha\beta}- k_\mathrm BTM_{\alpha\beta}|t|,~~~
\phi_{\alpha\beta}^\mathrm A(t)\approx-\frac{2k_\mathrm BT k^\mathrm o t}{\mathrm{tr}[M K^\mathrm e]}\det[M]\epsilon_{\alpha\beta}.
\label{app:OverTCF-gen-ST}
\end{align}
Similar to Eqs.~(\ref{OverTCF-e-ST}) and (\ref{OverTCF-o-ST}), the slopes of the symmetric and anti-symmetric parts are
proportional to the mobility tensor $M_{\alpha\beta}$ and the odd elasticity $k^\mathrm o$, respectively,
although the results are more general.

\end{widetext}


\end{document}